\documentclass[reprint, one column, aps, notitlepage]{revtex4-1}
\usepackage{amsmath}
\usepackage{physics}
\usepackage{graphicx}
\usepackage{gensymb}
\usepackage{subcaption}
\usepackage{float}
\graphicspath{ { /Users/diaza/Dropbox (MIT)/Lab/DPF/ } }

\begin{document}

\title{Status of Sterile Neutrino fits with Global Data}
\author{Alejandro Diaz}
\email{diaza@mit.edu}
\affiliation{Department of Physics, Massachusetts Institute of Technology, \\
Cambridge, MA 02139, USA}
\begin{abstract}

A number of short baseline neutrino oscillation experiments have observed an anomalous excess of neutrinos in the low energy range. This may hint towards the existence of additional neutrino mass splittings. If true, additional sterile (non-interacting) neutrino states above the current 3 neutrino model would be required. On the other hand, many parameters of the allowed space are limited by experiments that have seen no anomaly. We will introduce models which accommodate these additional neutrinos, and then discuss our work towards fitting these models to the available global high $\Delta m^{2}$ oscillation data. We will then present the latest results of these fits.

\end{abstract}

\maketitle

\large{Talk presented at the APS Division of Particles and Fields Meeting (DPF 2017), July 31-August 4, 2017, Fermilab. C170731}

\section{Introduction}

\subsection{Oscillation Formalism}

The current theory of neutrino oscillation proposes that the weak interaction eigenstates of neutrinos ($\nu_{e}$, $\nu_{\mu}$, $\nu_{\tau}$) are composed of a linear superposition of the mass eigenstates of the neutrinos, labeled as $\nu_{1}$, $\nu_{2}$, $\nu_{3}$. The relationship between these two bases is given by the Pontecovo-Maki-Nakagawa-Sakata (PMNS) Matrix:   
$$
\begin{pmatrix}
	\nu_{e} \\
	\nu_{\mu} \\
	\nu_{\tau} 
\end{pmatrix}
=
\begin{pmatrix}
	U_{e1} & U_{e2} & U_{e3} \\
	U_{\mu1} & U_{\mu2} & U_{\mu3} \\
	U_{\tau1} & U_{\tau2} & U_{\tau3} 
\end{pmatrix}	
\begin{pmatrix}
	\nu_{1} \\
	\nu_{2} \\
	\nu_{3} 
\end{pmatrix}.
$$

When a neutrino is produced by weak decay into one of its weak eigenstates and then travels through space, the composition of the neutrino state in the weak basis will oscillate. A detector, then, will have a non-zero probability of finding a neutrino of a different weak flavor than was originally produced. 

\begin{figure}
\includegraphics[width = 0.5\textwidth]{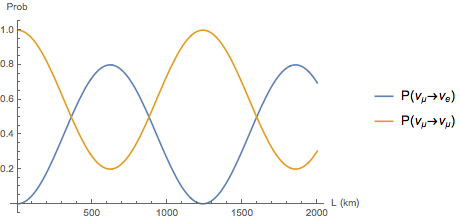}
\caption{This plot shows the probability of finding either $\nu_{e}$ or $\nu_{\mu}$ a distance L from where the $\nu_{\mu}$ was first created. The parameters used were $\Delta m^{2} = 0.002 \text{ eV}^{2}, \sin^{2}(2\theta) = 0.8, E_{\nu}=1 \text{ GeV}$.}
\label{fig:oscillation}
\end{figure}

This idea can be best illustrated if we assume a 2 neutrino model. In this case, there are only two weak and mass eigenstates for the neutrinos, and the mixing matrix is 2$\times$2 parameterized by a single parameter $\theta$:
$$
\begin{pmatrix}
	\nu_{e} \\
	\nu_{\mu}
\end{pmatrix}
=
\begin{pmatrix}
	U_{e1} & U_{e2} \\
	U_{\mu1} & U_{\mu2} \\
\end{pmatrix}	
\begin{pmatrix}
	\nu_{1} \\
	\nu_{2}
\end{pmatrix}
=
\begin{pmatrix}
	\text{cos }\theta & \text{sin }\theta \\
	-\text{sin }\theta & \text{cos }\theta
\end{pmatrix}	
\begin{pmatrix}
	\nu_{1} \\
	\nu_{2}
\end{pmatrix}.
$$
If we assume that a $\nu_{\mu}$ was produced with energy $E_{\nu}$, its initial state can be rewritten in the mass eigenstate basis as $\ket{\nu_{\mu}} = -\text{sin } \theta \ket{\nu_{1}} + \text{cos } \theta \ket{\nu_{2}}$. If we then let the state evolve with time as it propagates through space a distance L, the probability of finding a $\nu_{\mu}$ will be 
\begin{equation}\label{eq:3}
P(\nu_{\mu} \rightarrow \nu_{\mu}) = 1 - \text{sin}^{2}(2\theta)\text{sin}^{2}\left(1.27\frac{\Delta m^{2} \text{[eV}^{2}\text{]} \text{L[m]}}{E_{\nu}\text{[MeV]}}\right),
\end{equation}
where $\Delta m^{2}$ is the mass squared difference of the two mass eigenstates, L is the distance that the neutrino propagates, and $E_{\nu}$ is the energy of the neutrino. When a detector is used to find the same neutrino type as the one created, this is referred to as a \textit{disappearance} experiment. Conversely, the probability of finding a $\nu_{e}$, which was not present before, is 
\begin{equation}\label{eq:4}
P(\nu_{\mu} \rightarrow \nu_{e}) = \text{sin}^{2}(2\theta)\text{sin}^{2}\left(1.27\frac{\Delta m^{2} \text{[eV}^{2}\text{]} \text{L[m]}}{E_{\nu}\text{[MeV]}}\right).
\end{equation}
When a detector looks for a neutrino of a different type than the one initially produced, this is referred to as an \textit{appearance} experiment. As expected, the sum of these two probabilities is one in a two neutrino model. In Figure \ref{fig:oscillation}, we depict the changing probabilities of finding either $\nu_{e}$ or $\nu_{\mu}$ for some given parameters. 

In practice, we try to place the detector where we expect the first maximum (minimum) to be for an appearance (disappearance) experiment. This is where the clearest sign of oscillation will be. 

For a 3 neutrino model, we have a similar, albeit more complicated, probability oscillation equation
\begin{multline} \label{eq:1}
P(\nu_{\alpha}\rightarrow\nu_{\beta}) = \delta_{\alpha\beta} - 4 \sum_{j>i} \Re(U_{\alpha i}^{*} U_{\beta i} U_{\alpha j} U_{\beta j}^{*}) \sin^{2}\left(1.27\frac{\Delta m^{2} \text{[eV}^{2}\text{]} \text{L[m]}}{E_{\nu}\text{[MeV]}}\right) \\
+ 2 \sum_{j>i} \Im(U_{\alpha i}^{*} U_{\beta i} U_{\alpha j} U_{\beta j}^{*}) \sin\left(2.54\frac{\Delta m^{2} \text{[eV}^{2}\text{]} \text{L[m]}}{E_{\nu}\text{[MeV]}} \right), 
\end{multline}
where $\alpha$ and $\beta$ are the neutrino weak eigenstates, while $i$ and $j$ are the mass eigenstates. 

The PMNS matrix can also be written as a 3-rotation

$$
U_{PMNS} 
=
\begin{pmatrix}
	U_{e1} & U_{e2} & U_{e3} \\
	U_{\mu1} & U_{\mu2} & U_{\mu3} \\
	U_{\tau1} & U_{\tau2} & U_{\tau3} 
\end{pmatrix}	
=
\begin{pmatrix}
	1 & 0 & 0 \\
	0 & c_{23} & s_{23} \\
	0 & -s_{23} & c_{23} 
\end{pmatrix}
\begin{pmatrix}
	c_{13} & 0 & s_{13}e^{-i\delta} \\
	0 & 1 & 0 \\
	-s_{13}e^{i\delta} & 0 & c_{13}
\end{pmatrix}
\begin{pmatrix}
	c_{12} & s_{12} & 0 \\
	-s_{12} & c_{12} & 0 \\
	0 & 0 & 1
\end{pmatrix},
$$
where, for example $s_{12} \equiv \sin(\theta_{12})$. We find 3 mixing parameters, $\theta_{12}$, $\theta_{13}$, and $\theta_{23}$, and one CP violating phase $\delta$. We can see how $\delta$ results in CP violation if we go back to Eq. \eqref{eq:1}. If we wish to consider the oscillation of antineutrinos instead of neutrinos, then we would take the complex conjugates of the matrix elements, i.e. $U\rightarrow U^{*}$. Since the real part of a complex number does not change with complex conjugation, the term where the real part is taken in Eq. \eqref{eq:1} does not change. The term where we take the imaginary part, though, flips signs with complex conjugation. Thus, if $\delta \neq 0, 180$, then the matrix elements will have a complex component, and the oscillations between neutrino and antineutrinos will be different. This results in CP violation, since neutrinos and antineutrinos should oscillate the same if CP were conserved. 

\subsection{Important Values}

Here we summarize the best fit values for the above parameters, as given by \cite{nufit}. 

The magnitude of the PMNS matrix elements are found to be approximately:
$$
U_{PMNS} 
=
\begin{pmatrix}
	|U_{e1}| & |U_{e2}| & |U_{e3}| \\
	|U_{\mu1}| & |U_{\mu2}| & |U_{\mu3}| \\
	|U_{\tau1}| & |U_{\tau2}| & |U_{\tau3}| 
\end{pmatrix}	
\approx
\begin{pmatrix}
	0.82 & 0.55 & 0.15 \\
	0.38 & 0.57 & 0.70 \\
	0.39 & 0.59 & 0.69 
\end{pmatrix}
$$

The oscillation angles are found to be approximately: 
$$
\begin{matrix}
	\theta_{12} &\approx 34\degree \\
	\theta_{23} &\approx 42\degree \\
	\theta_{13} &\approx 8.5\degree
\end{matrix}	
\Longrightarrow
\begin{matrix}
	\sin^{2}(2\theta_{12}) &\approx 0.85 \\
	\sin^{2}(2\theta_{23}) &\approx 0.99 \\
	\sin^{2}(2\theta_{13}) &\approx 0.08
\end{matrix}
$$

We note that, unlike mixing in the quark sector, mixing in the neutrino is sector is very strong. The PMNS matrix is far from diagonal, and the mixing due to $\theta_{12}$ and $\theta_{23}$ is very strong. 

The CP violating term is found to be 
$$\delta= {261\degree}^{+51\degree}_{-59\degree}$$
We note that 360\degree \ (i.e. 0\degree) and 180\degree \ are both within 2$\sigma$ of the best fit value, so the likely value of $\delta$ still extends to values without CP violation. 

Most importantly, for the discussion that follows, we give the mass squared differences \cite{nufit}. 

\begin{equation} \label{eq:2}
\begin{aligned}
\Delta m_{21}^{2} &= (7.5 \pm 0.2) \times 10^{-5}\text{eV}^{2} \\
|\Delta m_{32}^{2}| &= (2.52 \pm 0.04) \times 10^{-3} \text{eV}^{2} 
\end{aligned}
\end{equation}

We note that these $\Delta m^{2}$ are \textit{small} compared to any other mass scale that we are familiar with in the Standard Model. 

\section{Anomalies}

We now discuss anomalies seen in a pair of oscillation experiments. 

\subsection{LSND}

The Liquid Scintillator Neutrino Detector (LSND) was an experiment at Los Alamos where a 798 MeV proton beam impinged upon a target to produce pions and muons that decayed at rest \cite{anomaly}.  The pions would decay as $\pi^{+}\rightarrow \mu^{+}\nu_{\mu}$, and then the muons would decay as $\mu^{+} \rightarrow e^{+}\nu_{e}\bar{\nu}_{\mu}$. LSND searched for $\bar{\nu}_{\mu} \rightarrow \bar{\nu}_{e}$ appearances. 

\begin{figure}
	\begin{subfigure}{0.5\textwidth}
		\includegraphics[width=0.9\linewidth]{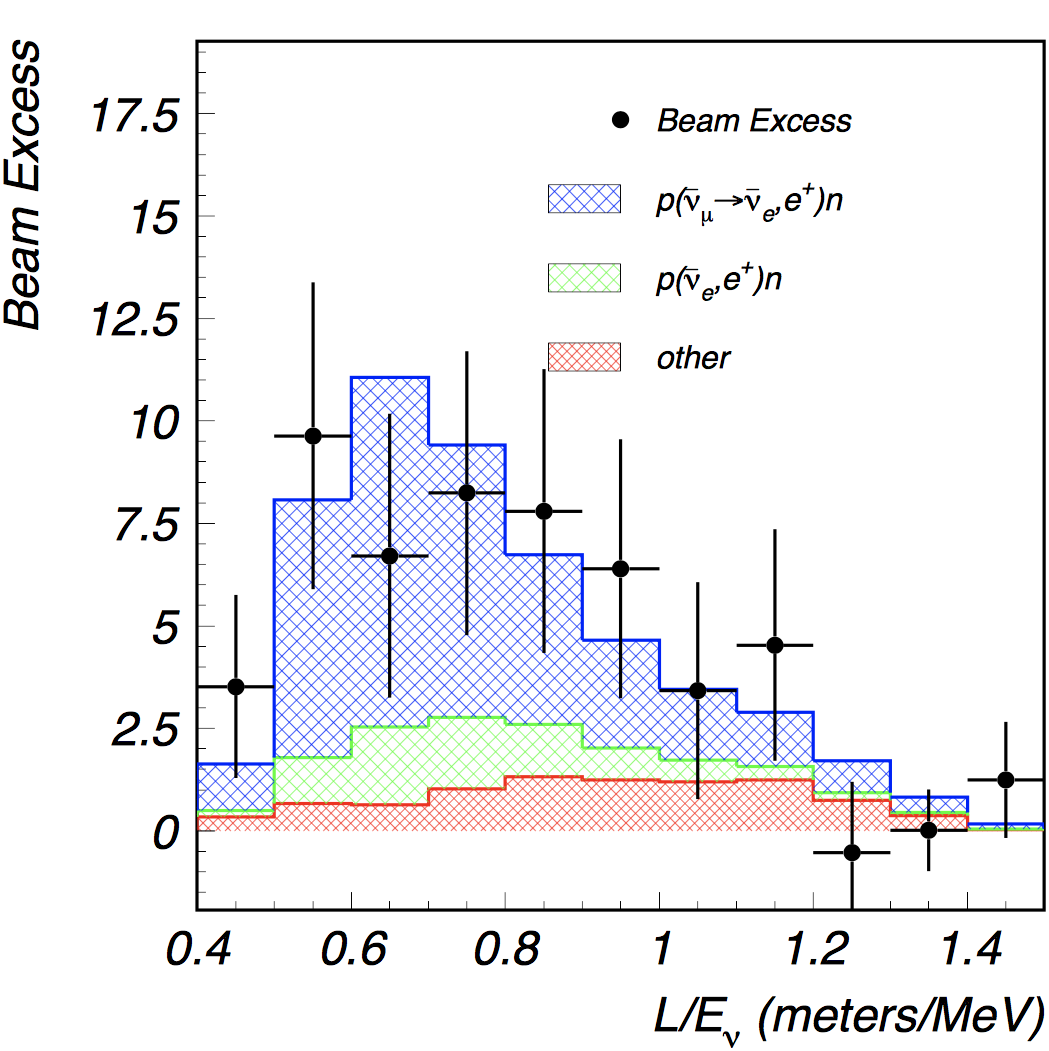}
		\caption{Excess of neutrino events detected in LSND above the expected background}
		\label{fig:lsndexcess}
	\end{subfigure}
	\begin{subfigure}{0.49\textwidth}
		\includegraphics[width=0.87\linewidth]{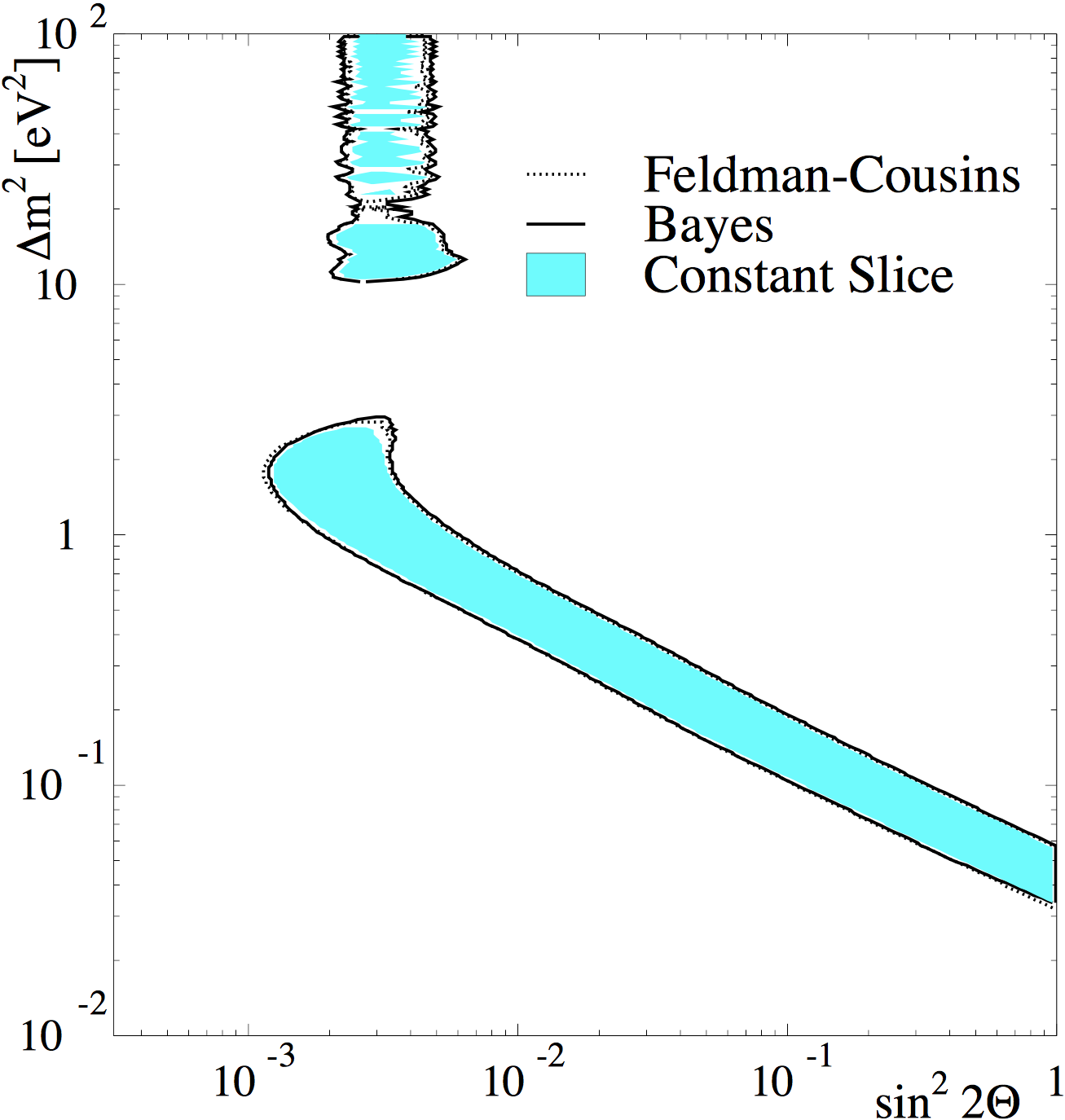}
		\caption{Best fit parameters of a neutrino, given the data from LSND}
		\label{fig:bestfit}
	\end{subfigure}
\end{figure}

LSND found an excess of $87.9 \pm 22.4 \pm 6.0$ $\bar{\nu}_{e}$ above the expected background \cite{lsnd}, corresponding to an oscillation probability of $(0.264 \pm 0.067 \pm 0.045)\%$. Figure \ref{fig:lsndexcess} shows this excess above the expected background.  

If this excess were treated as an oscillation between 2 neutrinos, its best fit parameters are found to be $\Delta m^{2} = 1.2 \text{eV}^{2}$ and $\sin^{2} 2\theta = 0.003$ (Figure \ref{fig:bestfit}). A $\Delta m^{2}$ this large is not consistent with the low $\Delta m^{2}$ in Eq. \eqref{eq:2} that are the currently accepted mass splittings. But, in order to have a third mass splitting, this would require a \textit{fourth} neutrino. 

\subsection{Can we have more neutrinos?}

We would like to ask if it's possible to have more than the 3 neutrinos that currently exist in the Standard Model. To try to answer this question, we consider the Z decay width \cite{thomson}. 

We would expect that the Z decay width is just the sum of the decay width of its decay channels, 
$$\Gamma_{Z} = 3\Gamma_{ll} + \Gamma_{\text{hadrons}} + N_{\nu}\Gamma_{\nu\nu},$$
where $\Gamma_{ll}$ is the decay width of the charged leptons (which we assume are all the same through lepton universality), $\Gamma_{\text{hadrons}}$ is the decay width to the hadrons, $\Gamma_{\nu\nu}$ is the decay width to the neutrinos, and $N_{\nu}$ is the number of neutrinos. 

$\Gamma_{Z}$, $\Gamma_{ll}$, and $\Gamma_{\text{hadrons}}$ are all known experimentally, and $\Gamma_{\nu\nu}$ can be predicted through SM predictions. If we then solve for $N_{\nu}$:
$$N_{\nu} = \frac{\Gamma_{Z} - 3\Gamma_{ll} - \Gamma_{\text{hadrons}}}{\Gamma_{\nu\nu}^{\text{SM}}},$$
we find that $N_{\nu} = 2.9840 \pm 0.0082$. This value of $N_{\nu}$ is consistent with the Z decaying to only 3 neutrinos. 

The most obvious way to bypass this requirement is if the fourth neutrino is more massive than half of the Z boson mass, making the decay kinematically forbidden. But given the upper bound of the neutrino masses \cite{pdg}, $m_{\nu} < 2 \text{eV}$, it is not possible that a fourth neutrino could reach up to a mass of $m_{Z}/2\approx 46\text{ GeV}$ with the $\Delta m^{2}$ found by LSND. This, then, could not explain the mass splitting seen in LSND. 

Another hypothesis, which we'll explore through the rest of these proceedings, is that this new neutrino does not interact weakly. This hypothetical particle is called a \textit{sterile} neutrino. 

\subsection{Adding a neutrino}

To add more neutrinos to our model, we simply add a new row and column to our old PMNS matrix. We refer to one additional neutrino as the 3+1 model. 
$$
\begin{pmatrix}
	\nu_{e} \\
	\nu_{\mu} \\
	\nu_{\tau} 
\end{pmatrix}
=
\begin{pmatrix}
	U_{e1} & U_{e2} & U_{e3} \\
	U_{\mu1} & U_{\mu2} & U_{\mu3} \\
	U_{\tau1} & U_{\tau2} & U_{\tau3} 
\end{pmatrix}	
\begin{pmatrix}
	\nu_{1} \\
	\nu_{2} \\
	\nu_{3} 
\end{pmatrix}
\Rightarrow
\begin{pmatrix}
	\nu_{e} \\
	\nu_{\mu} \\
	\nu_{\tau} \\
	\nu_{s}
\end{pmatrix}
=
\begin{pmatrix}
	U_{e1} & U_{e2} & U_{e3}  & U_{e4}\\
	U_{\mu1} & U_{\mu2} & U_{\mu3} & U_{\mu4}\\
	U_{\tau1} & U_{\tau2} & U_{\tau3} & U_{\tau4}\\
	U_{s1} & U_{s2} & U_{s3} & U_{s4}
\end{pmatrix}	
\begin{pmatrix}
	\nu_{1} \\
	\nu_{2} \\
	\nu_{3} \\
	\nu_{4}
\end{pmatrix}
$$

This new form adds 7 matrix elements, 1 mass splitting, and 2 CP violating complex phases \cite{gabriel1}. The four $U_{si}$ matrix elements cannot be constrained because we cannot directly observe sterile neutrinos.

Additionally, since the observed mass splitting in LSND, $\Delta m^{2}_{41}$, is much larger than the two SM mass splittings,  we will use the short-baseline approximation, where we assume that $\Delta m^{2}_{21}$ and $\Delta m^{2}_{32}$ are both 0 and that the 3 SM neutrino states are degenerate. Short-baseline refers to experiments with $L/E \ll 1$, so that when $\Delta m^{2} \ll 1$, 
$$\sin^{2}\left(1.27 \frac{\Delta m^{2} \text{[eV}^{2}\text{]} \text{L[m]}}{E_{\nu}\text{[MeV]}}\right) \approx 0 $$
and terms due to these $\Delta m^{2}$ can be negleted. 

Assuming this, the probability equation \eqref{eq:1} (which is valid for an arbitrary number of neutrinos), becomes 
\begin{equation}\label{eq:5}
P(\nu_{\alpha} \rightarrow \nu_{\beta}) = \delta_{\alpha\beta} - 4(\delta_{\alpha\beta}-U_{\alpha4}^{*}U_{\beta4})U_{\alpha4}U_{\beta4}^{*})\sin^{2}\left(1.27 \frac{\Delta m^{2} \text{[eV}^{2}\text{]} \text{L[m]}}{E_{\nu}\text{[MeV]}}\right).
\end{equation}
Note that since eq \eqref{eq:5} is real, then there is no CP violation in this limit. 

We can define
$$\sin^{2} 2\theta_{\alpha\beta} = |4(\delta_{\alpha\beta}-U_{\alpha4}^{*}U_{\beta4})U_{\alpha4}U_{\beta4}^{*}|$$

If we let $\alpha = \beta$, then we get
$$P(\nu_{\alpha} \rightarrow \nu_{\alpha}) = \delta_{\alpha\beta} - \sin^{2} 2\theta_{\alpha\beta} \sin^{2}\left(1.27 \frac{\Delta m^{2} \text{[eV}^{2}\text{]} \text{L[m]}}{E_{\nu}\text{[MeV]}}\right).$$
And if we let $\alpha \ne \beta$, then 
$$P(\nu_{\alpha} \rightarrow \nu_{\beta}) = \sin^{2} 2\theta_{\alpha\beta} \sin^{2}\left(1.27 \frac{\Delta m^{2} \text{[eV}^{2}\text{]} \text{L[m]}}{E_{\nu}\text{[MeV]}}\right).$$

We notice that, in this limit, the form of the oscillation equations for a one sterile neutrino model are identical to the 2 neutrino oscillation probabilities given in Equations \eqref{eq:3} and \eqref{eq:4}. We thus find that the 3+1 model in the short-baseline limit is similar to a 2 neutrino model. 

\subsection{Adding More Neutrinos}

If we want to add a second neutrino (a 3+2 model), we simply add another row and another column to the 3+1 model.
$$
\begin{pmatrix}
	\nu_{e} \\
	\nu_{\mu} \\
	\nu_{\tau} \\
	\nu_{s_{1}} \\
	\nu_{s_{2}}
\end{pmatrix}
=
\begin{pmatrix}
	U_{e1} & U_{e2} & U_{e3}  & U_{e4} & U_{e5}\\
	U_{\mu1} & U_{\mu2} & U_{\mu3} & U_{\mu4} & U_{\mu5}\\
	U_{\tau1} & U_{\tau2} & U_{\tau3} & U_{\tau4} & U_{\tau5}\\
	U_{s_{1}1} & U_{s_{1}2} & U_{s_{1}3} & U_{s_{1}4} & U_{s_{1}5}\\
	U_{s_{2}1} & U_{s_{2}2} & U_{s_{2}3} & U_{s_{2}4} & U_{s_{2}5}
\end{pmatrix}	
\begin{pmatrix}
	\nu_{1} \\
	\nu_{2} \\
	\nu_{3} \\
	\nu_{4} \\
	\nu_{5}
\end{pmatrix}
$$

The 3+2 model adds 6 \textit{constrainable} matrix elements $U_{li}$, where $l$ are the SM leptons, and $i \in (4,5)$. In the short-baseline approximation, we also have two mass splittings $\Delta m_{41}^{2}$ and $\Delta m_{51}^{2}$, and a CP violating phase $\Phi_{41}$ is introduced.

If we would like to continue to add neutrinos to the model, we just add additional rows and columns as needed. 

\subsection{MiniBooNE}

The MiniBooNE experiment at Fermilab collided 8 GeV kinetic energy protons with a target to produce $\pi$'s and K's that would then decay in flight. The mesons would decay into neutrinos which were picked up by a detector 541 m downstream. While the energy and baseline of MiniBooNE differed from LSND, the L/E were similar, so that MiniBooNE was sensitive to the same parameter space as LSND. 

A magnetic focusing horn downstream from the target was able to focus either the $\pi^{+}$'s and K$^{+}$'s, or the $\pi^{-}$ and K$^{-}$, and thus alternate between neutrino and antineutrino mode, searching for either $\nu_{\mu}\rightarrow \nu_{e}$ or $\bar{\nu}_{\mu} \rightarrow \bar{\nu}_{e}$ oscillations. 

\begin{figure}
\includegraphics[width = 0.5\textwidth]{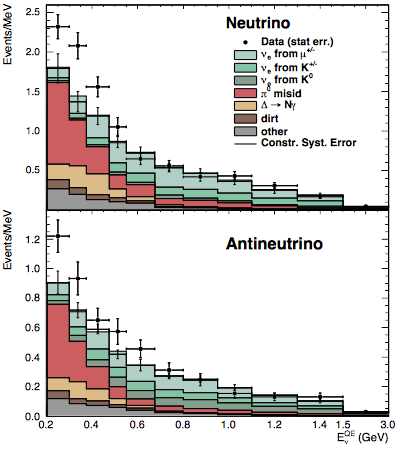}
\caption{The number of neutrino events seen in MiniBooNE, separated by neutrino and antineutrino mode. The excess is primarily seen under 475 MeV in neutrino mode, while the excess is seen over a wider range in antineutrino mode. }
\label{fig:minibooneexcess}
\end{figure}

MiniBooNE found an excess of events above background in both neutrino and antineutrino mode \cite{miniboone}. In neutrino mode, an excess of $160.0 \pm 47.8$ $\nu_{e}$ events were found above background, primarily under 475 MeV. In antineutrino mode, an excess of $78.4 \pm 28.5$ $\bar{\nu}_{e}$ events were found above background, both above and below 475 MeV. The excess of LSND, which only ran in antineutrino mode, was found in a similar L/E range as MiniBooNE's antineutrino range. The excess can be seen in Figure \ref{fig:minibooneexcess}.

In Figure \ref{fig:minibooneexcess}, we notice that a large fraction of the background comes from $\pi^{0}$ misidentification. A major contributor to this is that MiniBooNE could not differentiate between photons and electrons. A $\pi^{0}$ decays into 2 photons, while a neutrino interaction emits only one electron. If the detector picks up two particles, it's tagged as a $\pi^{0}$ decay and can be cut out. But, for example, if a $\pi^{0}$ decayed back to back photons, the forward boosted photon would leave a signal while the backward boosted photon could have very low energy and not be observed. So the detector only sees one particle, and misidentifies the event as a neutrino event. The MicroBooNE experiment intends to address this issue.  

\begin{figure}
\includegraphics[width = 0.35\textwidth]{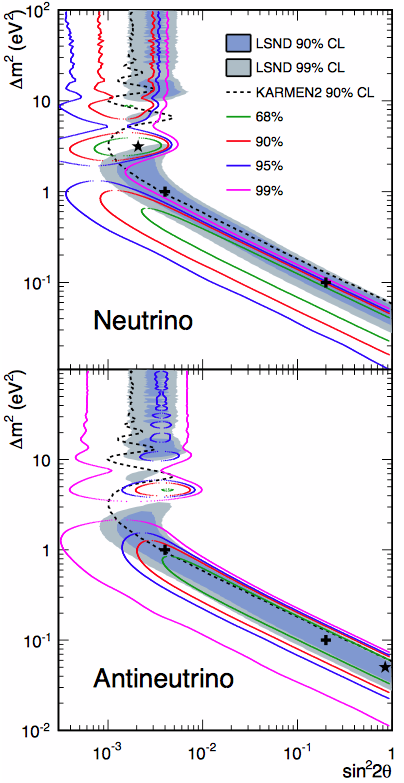}
\caption{The best fit of the neutrino and antineutrino MiniBooNE excess, overlaid with the best fit of the LSND excess.}
\label{fig:bestfitminiboone}
\end{figure}

As can be seen in Figure \ref{fig:minibooneexcess}, the neutrino and antineutrino excess occur in different energy ranges. When we compare the regions of best fit in Figure \ref{fig:bestfitminiboone}, we can see that the neutrino excess is marginally compatible with the LSND excess. The antineutrino excess, on the other hand, has a best fit parameter space that overlaps the LSND parameter space much more. This difference in compatibility between the two data sets  could indicate that neutrinos and antineutrinos oscillate differently, and could then be evidence of CP violation. As noted earlier, CP violation does not appear in the 3+1 sterile neutrino model. Thus, the MiniBooNE excess could point towards a model with more than 1 sterile neutrino. 

\section{Overview of Experiments}

\begin{figure}
\centering
\Large{\textbf{$\nu_{\mu}\rightarrow \nu_{e}$ and $\bar{\nu}_{\mu}\rightarrow\bar{\nu}_{e}$}}\par\medskip
\includegraphics[width = 0.5\textwidth]{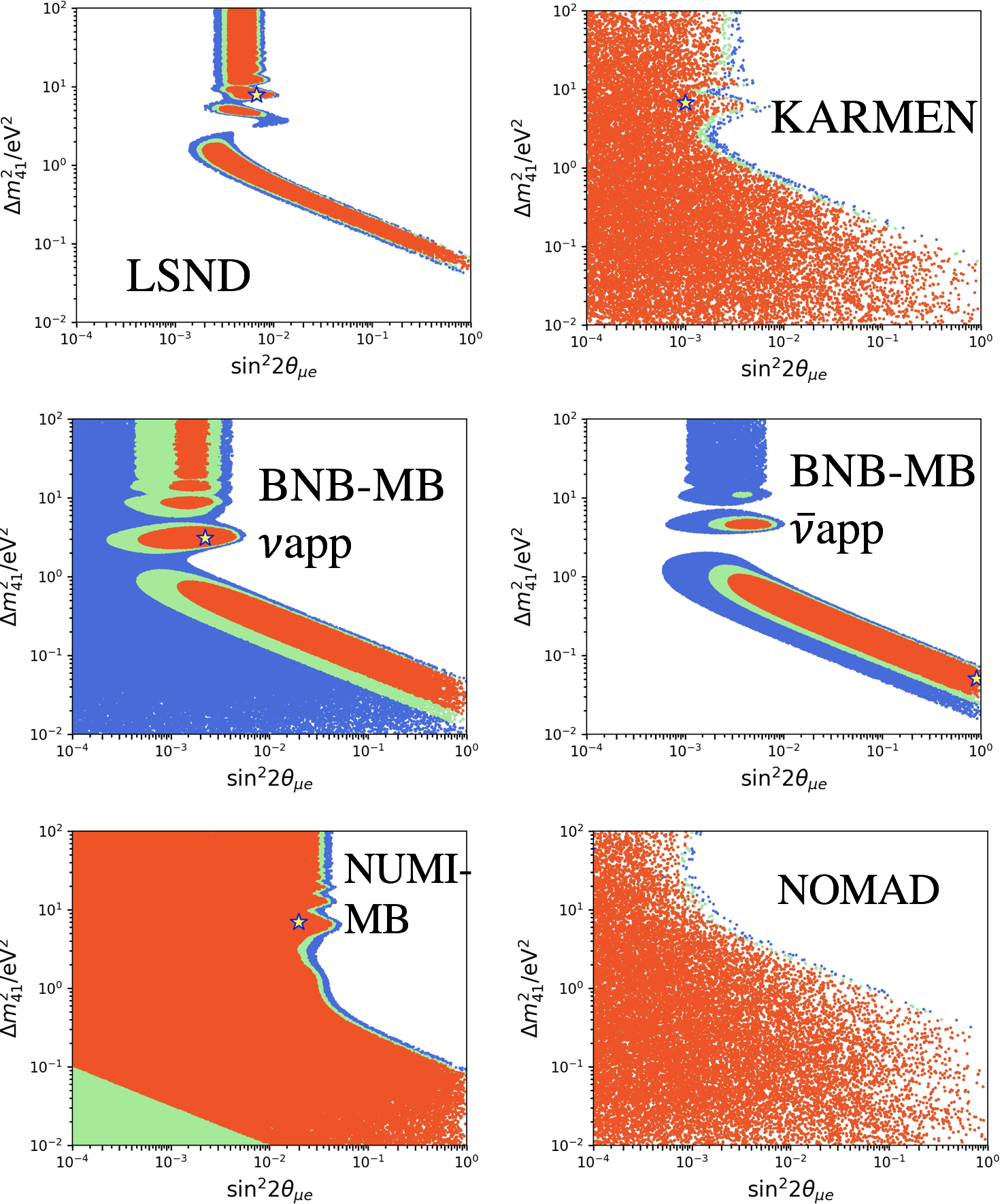}
\caption{Best fit parameter space for several $\nu_{\mu}\rightarrow \nu_{e}$ and $\bar{\nu}_{\mu}\rightarrow\bar{\nu}_{e}$ experiments}
\label{fig:eappearance}
\end{figure}

\begin{figure}
\centering
\Large{\textbf{$\nu_{\mu}\rightarrow \nu_{\mu}$ and $\bar{\nu}_{\mu}\rightarrow\bar{\nu}_{\mu}$}}\par\medskip
\includegraphics[width = 0.5\textwidth]{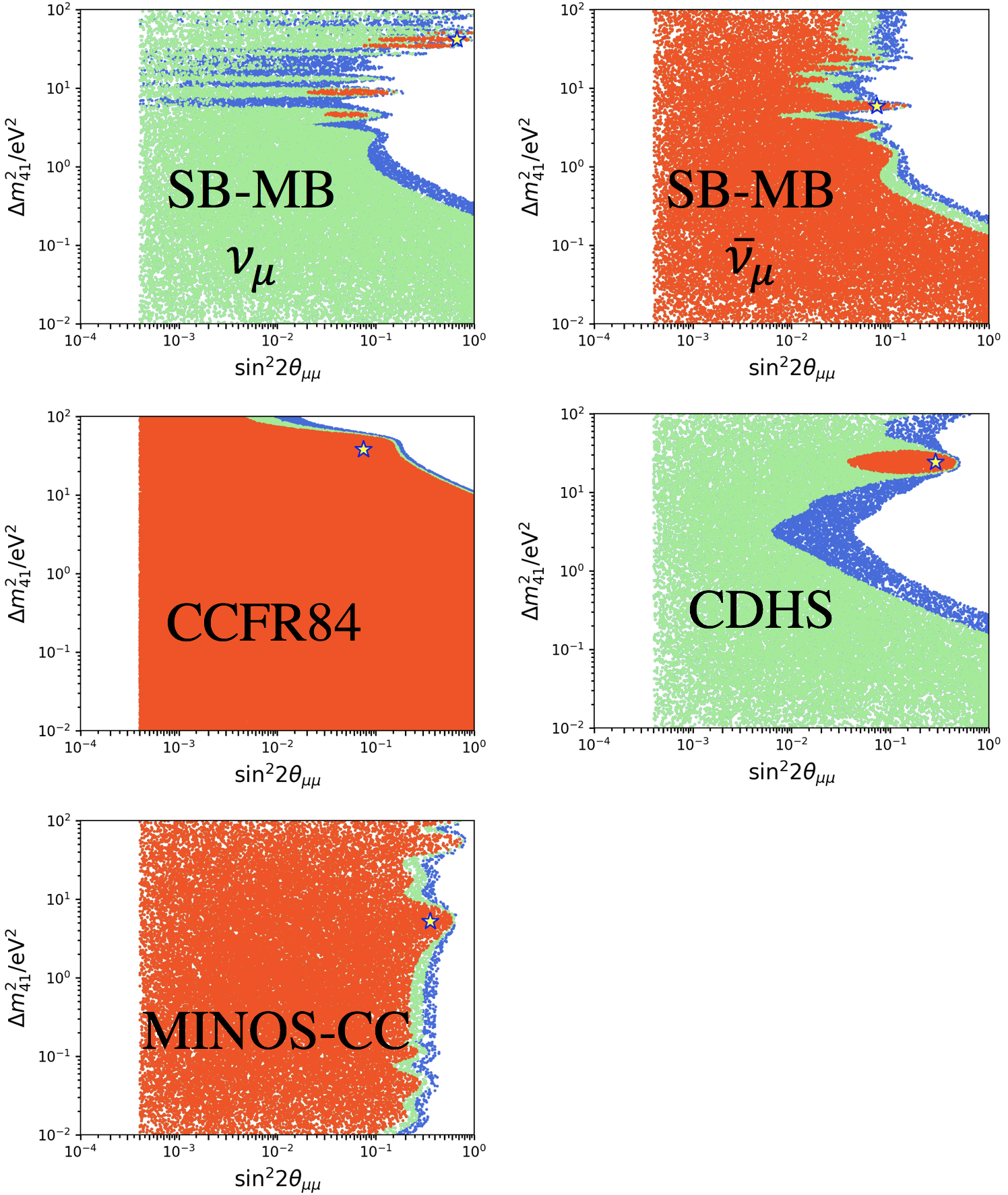}
\caption{Best fit parameter space for several $\nu_{\mu}\rightarrow \nu_{\mu}$ and $\bar{\nu}_{\mu}\rightarrow\bar{\nu}_{\mu}$ experiments}
\label{fig:mudisappearance}
\end{figure}

\begin{figure}
\centering
\Large{\textbf{$\nu_{e}\rightarrow \nu_{e}$ and $\bar{\nu}_{e}\rightarrow\bar{\nu}_{e}$}}\par\medskip
\includegraphics[width = 0.5\textwidth]{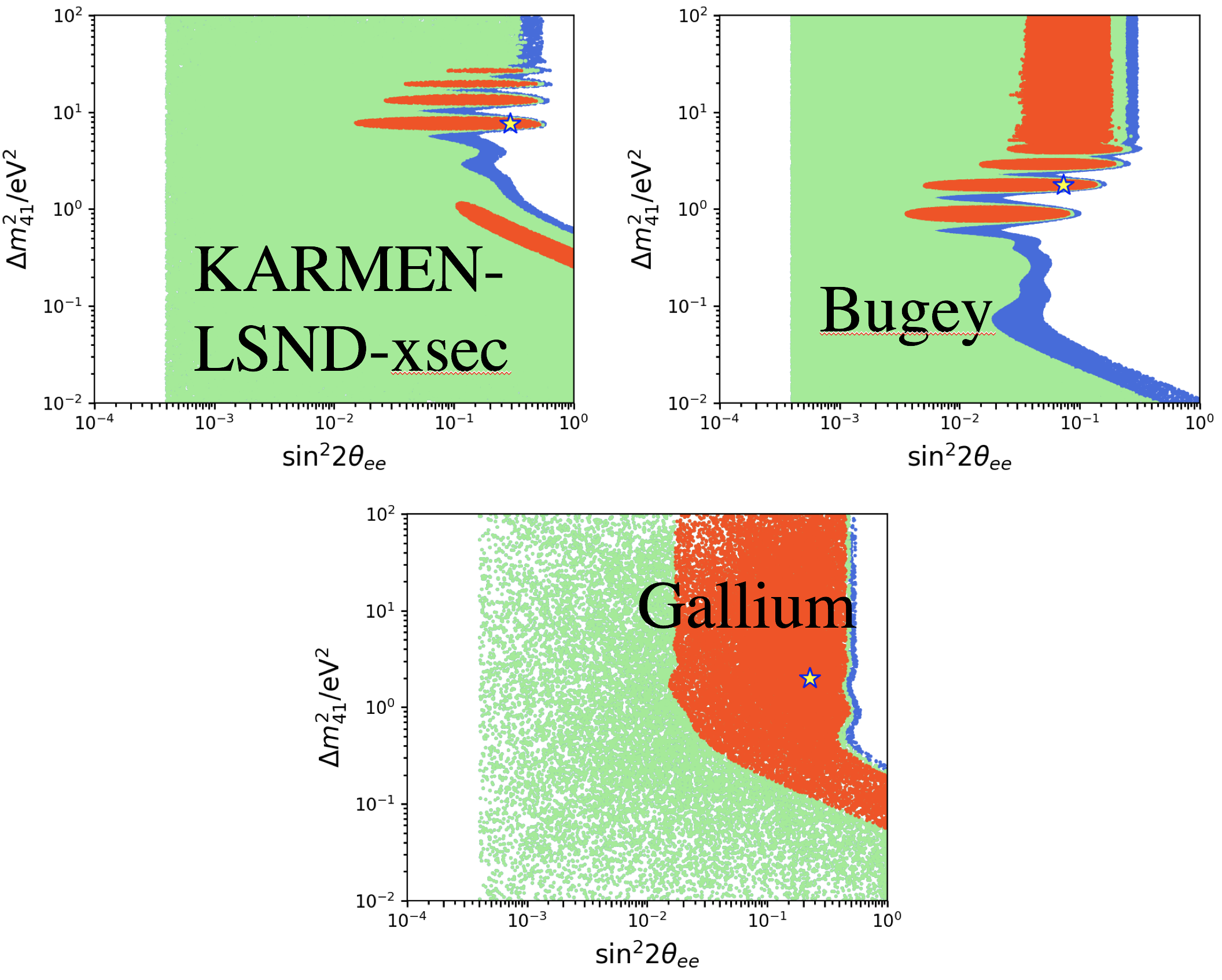}
\caption{Best fit parameter space for several $\nu_{e}\rightarrow \nu_{e}$ and $\bar{\nu}_{e}\rightarrow\bar{\nu}_{e}$ experiments}
\label{fig:edisappearance}
\end{figure}

While LSND and MiniBooNE are frequently cited as the motivation for the search of sterile neutrinos, there have been several other short-baseline experiments that have either seen or not seen a signal for sterile neutrinos. In Figures \ref{fig:eappearance}-\ref{fig:edisappearance}, we display fits to a 3+1 model using the oscillation data observed in each experiment used in our global fits. The red region is at 90\% confidence level, the green is at 2$\sigma$, and the blue is at 99\%. 

The definition for what constitutes a signal in the community is subjective, since we can arbitrarily choose a significance to classify as a ``signal." The classification of a signal in the community can also be inconsistent. For example, Bugey and Gallium in Figure \ref{fig:edisappearance} are considered to have seen a signal while only being significant at the 90\%, while CDHS in Figure \ref{fig:mudisappearance} is not considered a signal despite being significant at the same confidence level.

In a 3+1 model with the short-baseline approximation, we fit for three parameters, $\Delta m_{41}^{2}$, $|U_{e4}|$, and $|U_{\mu4}|$. Due to the high production threshold for $\tau$, we cannot directly constrain $|U_{\tau4}|$. Also, $|U_{e4}|$ and $|U_{\mu4}|$ can be rewritten as $\sin^{2} 2\theta_{\alpha\beta}$, where the $\alpha$ and $\beta$ used correspond on the type of oscillation the experiment is observing. 

In Figure \ref{fig:eappearance}, we show the fits for the $\nu_{\mu}\rightarrow \nu_{e}$ and $\bar{\nu}_{\mu}\rightarrow\bar{\nu}_{e}$ appearance experiments. In the middle plots, ``BNB-MB" refers to the MiniBooNE results using the Booster Neutrino Beam. ``NUMI-MB" stands for data from the MiniBooNE detector having used the NUMI beam line. The above analyses of MiniBooNE was only with the BNB line. 

We see that both LSND and MiniBooNE (using the BNB beam) in antineutrino mode see a signal at all confidence levels, while MiniBooNE (using BNB) in neutrino mode only sees a signal at the 90\% and $2\sigma$ level. KARMEN, NOMAD, and MiniBooNE using the NUMI beam do not see a signal.

In Figure \ref{fig:mudisappearance}, we show the fits for the $\nu_{\mu}\rightarrow \nu_{\mu}$ and $\bar{\nu}_{\mu}\rightarrow\bar{\nu}_{\mu}$ disappearance experiments. In the top row,``SB-MB" refers to a joint SciBooNE-MiniBooNE analyses. 

These are all regarded by the community to have not seen a signal. But, as we can see, both the SciBooNE-MiniBooNE joint neutrino mode analyses and CDHS see a signal at the 90\% confidence level. The SciBooNE-MiniBooNE joint antineutrino analysis, CCFR84, and MINOS all see no signal.

In Figure \ref{fig:edisappearance}, we show the fits for the $\nu_{e}\rightarrow \nu_{e}$ and $\bar{\nu}_{e}\rightarrow\bar{\nu}_{e}$ disappearance experiments. Of these, only Bugey and the Gallium experiments are considered by the community to have seen a signal, but we can see that they are significant only at 90\% and $2\sigma$. Also, the KARMEN-LSND-xsec analyses sees a signal at the same confidence level, but it is not generally considered a signal. 

\section{Global Fits}

Here, we present the sterile neutrino global fit analysis \cite{gabriel1}, where the data from all the above experiments are combined to find the best fit point in parameter space for a sterile neutrino. 

\subsection{3+1 Model}

\begin{figure}
\includegraphics[width=0.4\textwidth]{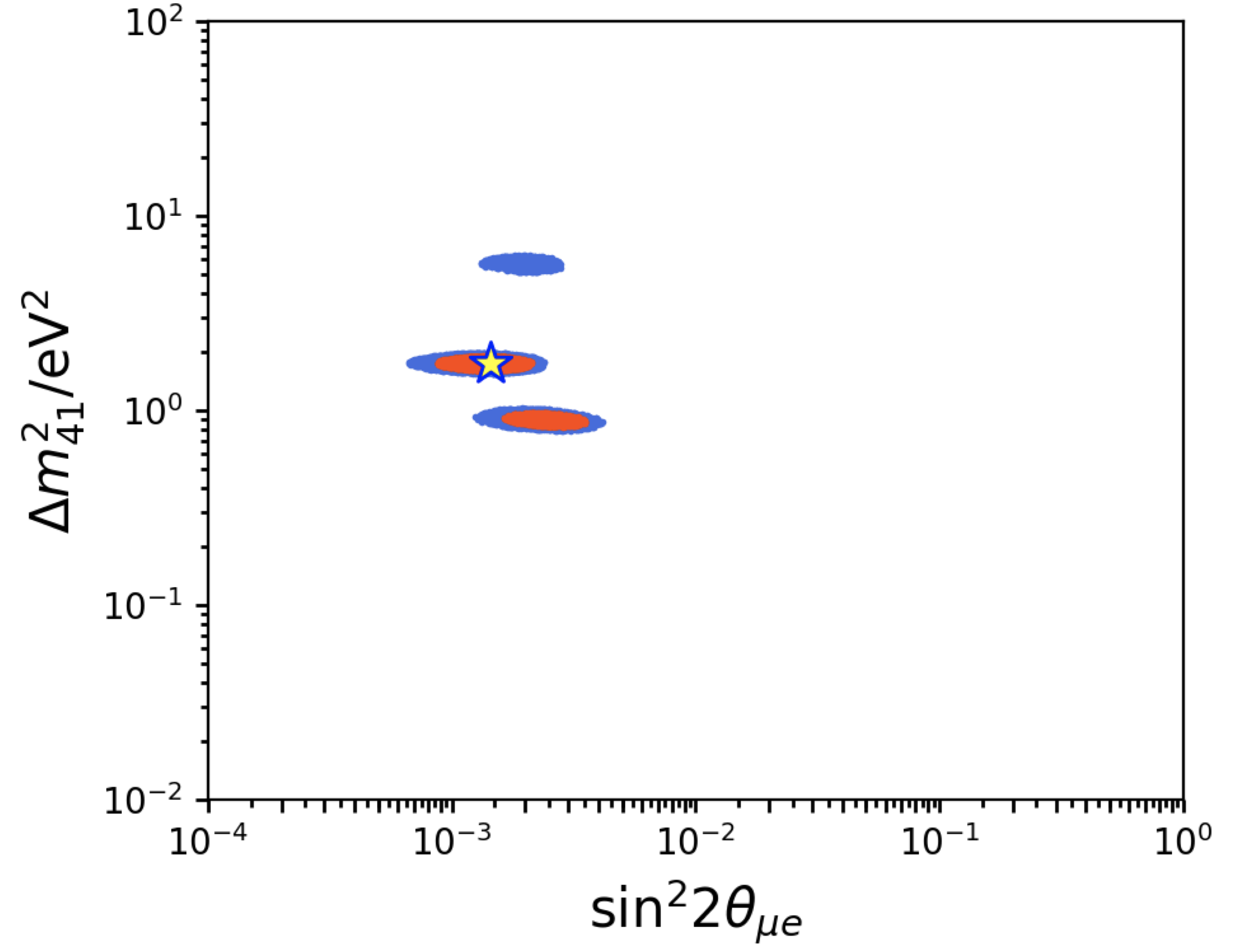}
\caption{The best fit regions for the parameters $\Delta m_{41}^{2}$ and $\sin^{2} 2\theta_{\mu e}$ in a 3+1 model}
\label{fig:global}
\end{figure}

We present the global fits assuming only one sterile neutrino. In a 3+1 model with the short-baseline approximation, the three relevant parameters to fit are $\Delta m_{41}^{2}$, $|U_{e4}|$, and $|U_{\mu4}|$. We show the best fit parameter space in Figure \ref{fig:global}. 
The best fit parameters we find are 
\begin{equation*}
\begin{aligned}
\Delta m_{41}^{2} &= 1.75 \text{eV}^{2} \\
|U_{e4}| &= 0.163 \\
|U_{\mu4}| &= 0.117
\end{aligned}
\end{equation*} 
We can rewrite the matrix elements as a mixing angle, giving:
$$
\sin^{2} 2\theta_{\mu e} = 1.45\times10^{-3}
$$

The $\chi^{2}$ at the best fit point and with the null hypothesis are
\begin{equation*}
\begin{aligned}
\chi_{\text{bf}}^{2} &= 306.81 \ (318 \text{ dof}) \\
\chi_{\text{null}}^{2} &= 359.15 \ (315 \text{ dof})
\end{aligned}
\end{equation*}
This gives us a $\chi^{2}$ difference of 
$$\Delta\chi_{\text{null-bf}}^{2} =  52.34 \ (3 \ \Delta\text{dof} )$$
By adding only 3 parameters to our model, we've decreased the $\chi^{2}$ by over 50. This indicates that the data seen in the above experiments much prefer the 3+1 sterile neutrino model over the null 3 neutrino model. 

\begin{figure}
\includegraphics[width=0.75\textwidth]{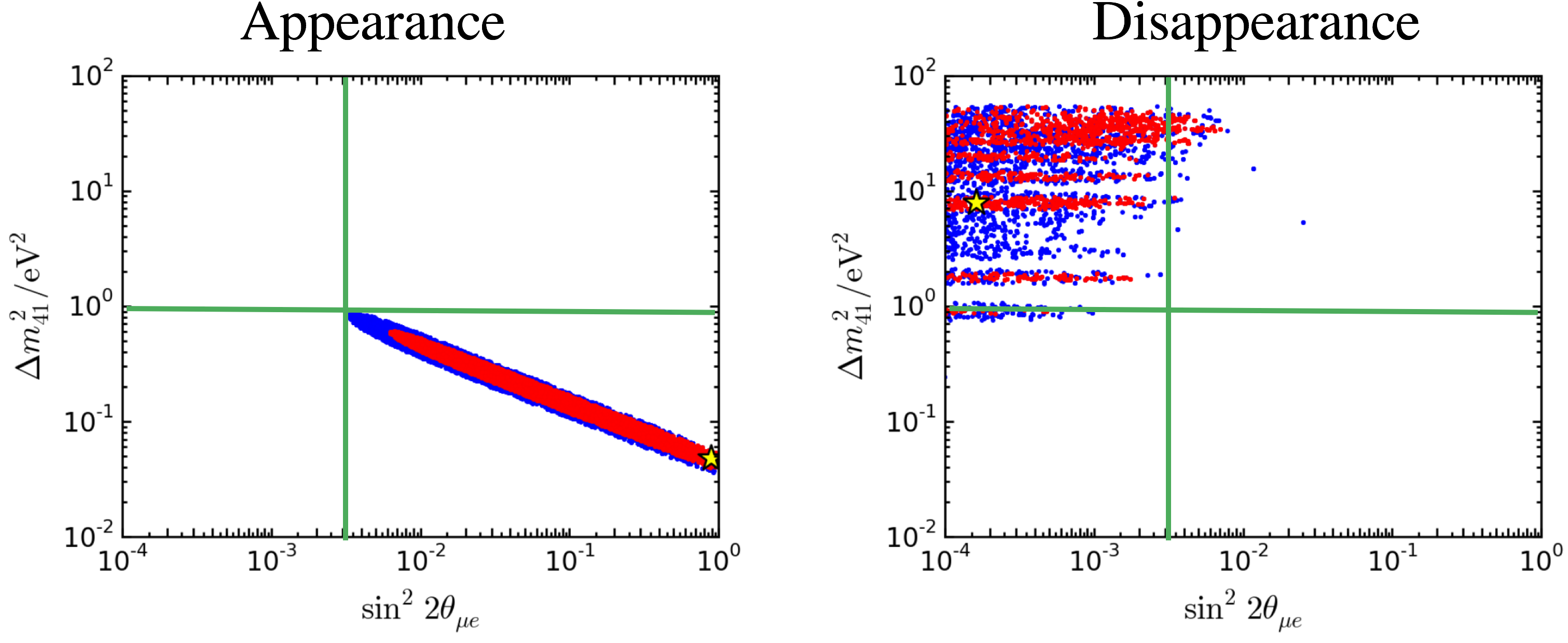}
\caption{The best fit confidence levels when we split the experiments between appearance and disappearance experiments. We find significant tension when we do this. The confidence levels do not overlap at all the 99\% and 90\% level. The green lines are included only to help show the tension.}
\label{fig:appdistension}
\end{figure}

\begin{figure}
\includegraphics[width=0.75\textwidth]{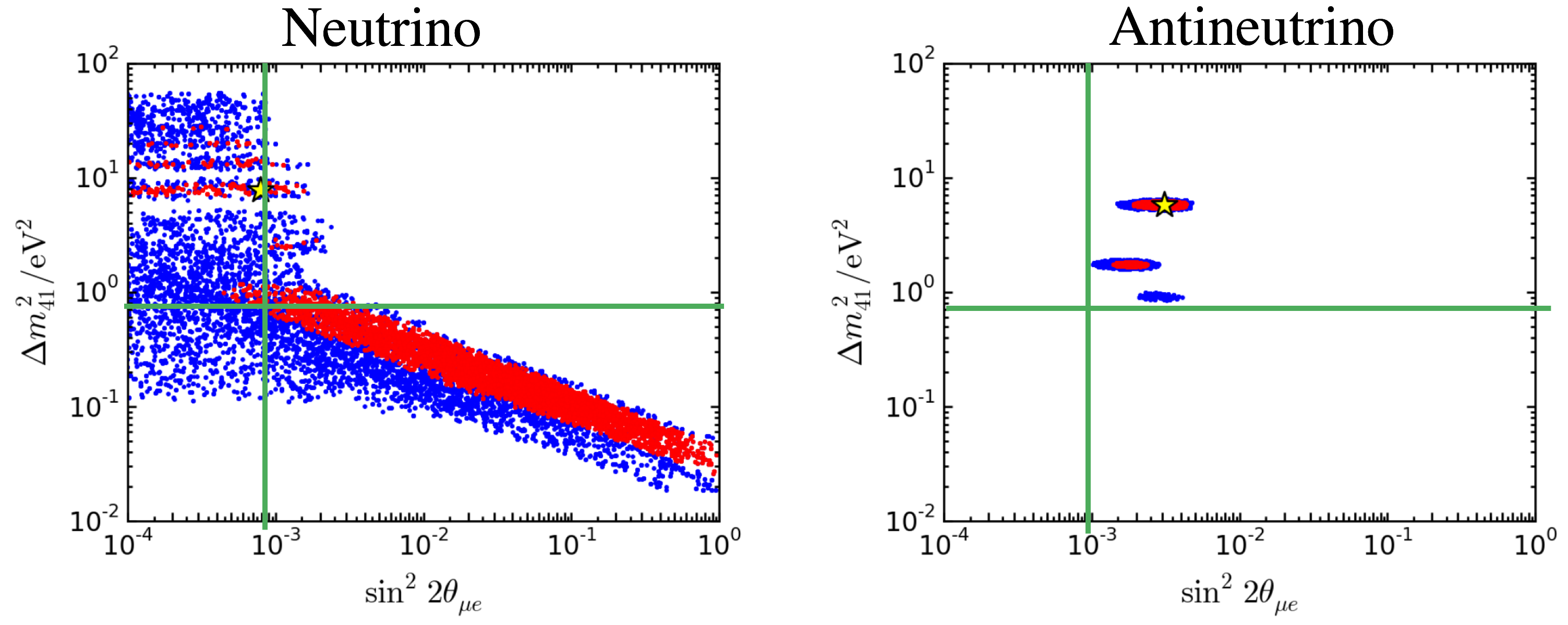}
\caption{The best fit confidence levels when we split the experiments between neutrino and antineutrino experiments. We find significant tension between the two datasets, with very little overlap in the 99\% and 90\% confidence level. The green lines are included only to help show the tension.}
\label{fig:neuantitension}
\end{figure}

Despite the above result, we do find significant tension within the datasets. If we split the data between appearance and disappearance experiments, we see in Figure \ref{fig:appdistension} that the 99\% and 90\% confidence do not overlap at all (though they do overlap if we go to higher confidence levels). We thus find tension between our appearance and disappearance experiments. 
Similarily, we see in Figure \ref{fig:neuantitension} that if we separate our data between neutrino and antineutrino data,  we again see very little overlap between the confidence levels. 

Because we find tension between neutrino and antineutrino oscillations, we might be inclined to try a 3+2 model. As we noted earlier, a CP violating phase is introduced in the 3+2 model that is not present in the 3+1 model. This accommodates differences between neutrino and antineutrino oscillations. 

\subsection{3+2 Model}

We present here the global best fit assuming two sterile neutrinos. As mentioned earlier, a 3+2 model with the short-baseline approximation includes a CP violating term, possibly resolving the tension between neutrino and antineutrino oscillation data found with the 3+1 global fit. In the 3+2 model, we fit for 7 parameters, $\Delta m_{41}^{2}$, $\Delta m_{51}^{2}$, $|U_{e4}|$, $|U_{\mu4}|$, $|U_{e5}|$, $|U_{\mu5}|$, and $\Phi_{41}$. Due to the high production threshold for $\tau$, we cannot directly constrain $|U_{\tau4}|$ and $|U_{\tau5}|$.

\begin{figure}
\includegraphics[width = 0.75\textwidth]{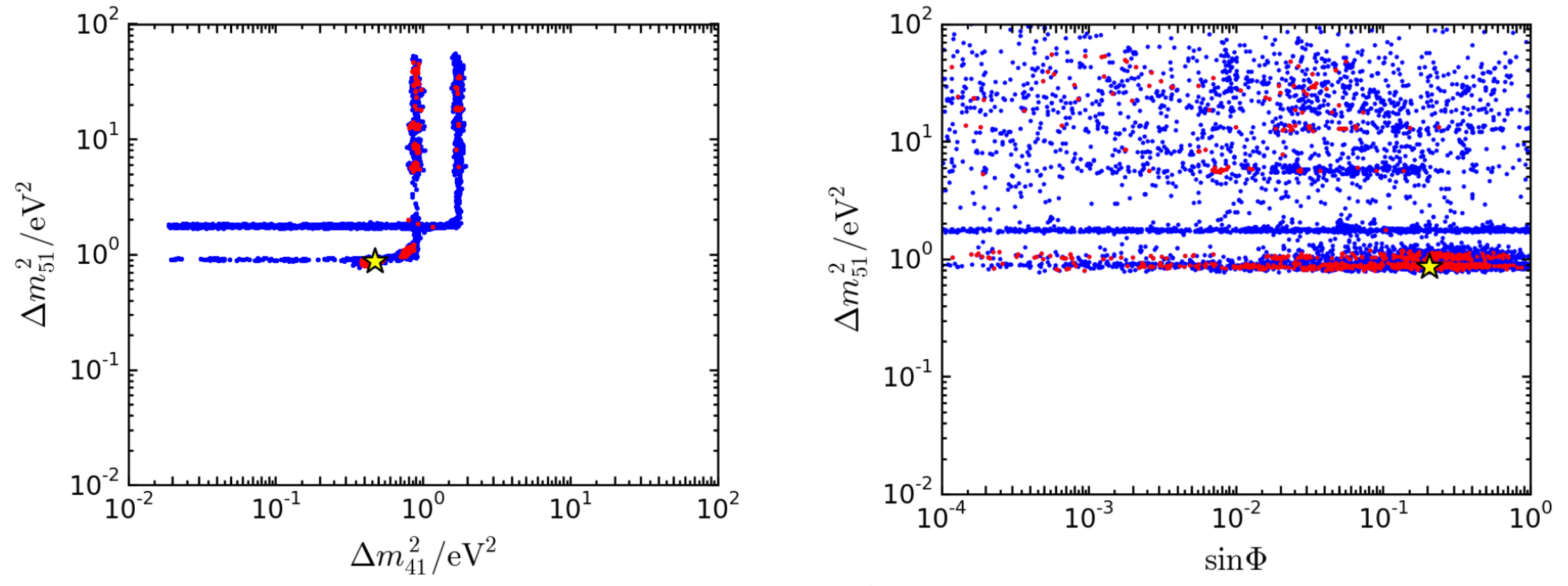}
\caption{Best fit confidence levels of a 3+2 model. On the left, we are plotting the $\Delta m_{41}^{2}$ vs. $\Delta m_{51}^{2}$. On the right, we plot $\sin \Phi$ vs. $\Delta m_{51}^{2}$}
\label{fig:3plus2}
\end{figure}

In Figure \ref{fig:3plus2}, we show a pair of best fit plots. On the left, we are plotting the $\Delta m_{41}^{2}$ vs. $\Delta m_{51}^{2}$. On the right, we plot $\sin \Phi$ vs. $\Delta m_{51}^{2}$. Note that we can only plot 2 out of the 7  parameters that are fitted in the 3+2 model at a time. We  end up with a $\chi^{2}$ difference of 
$$\Delta\chi_{\text{null-bf}}^{2} = 56.99 \ (7 \ \Delta\text{dof})$$
Compared to the 3+1 model, we improve our $\Delta \chi^{2}$ by ~4, while increasing the degrees of freedom by 4. Thus, this model doesn't do much towards improving our fits, despite introducing CP violation. This is illustrated in the plot on the right of Figure \ref{fig:3plus2}: the likely space for the CP violating term $\sin\Phi$ does not concentrate anywhere, and would be likely to lie at any value. 

\subsection{Summary}

To summarize what we have found, the introduction of a sterile neutrino with a 3+1 model significantly improves the fits with the oscillation data compared to the null 3 neutrino model. But the fits show a clear and strong tension when the data sets are separated between appearance and disappearance experiments, as well as separating between neutrino and antineutrino data. Introducing a CP violating phase by going to a 3+2 model doesn't improve the fit by much. 

\section{Future}

\subsection{MicroBooNE}

\begin{figure}
\includegraphics[width=0.35\textwidth]{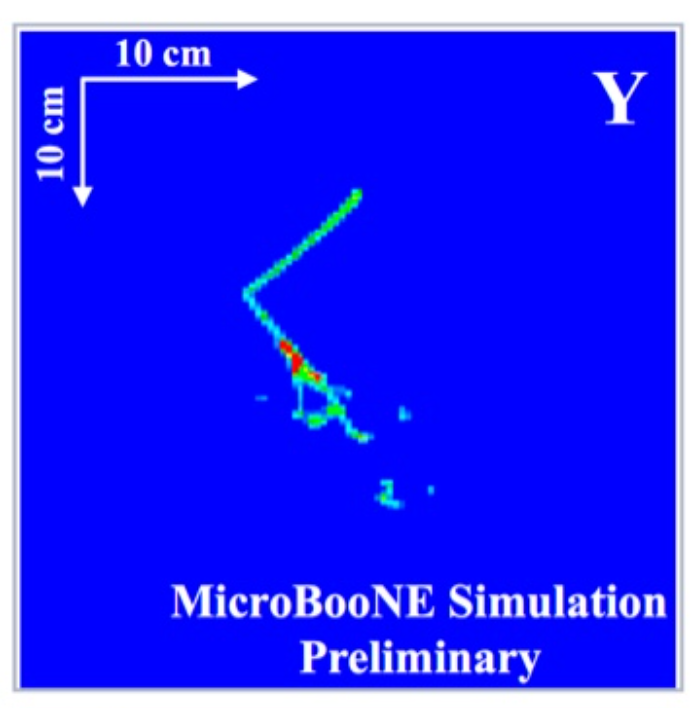}
\caption{A simulated event in MicroBooNE, where an electron and proton track can be seen with high resolution.}
\label{fig:microboone}
\end{figure}

As mentioned earlier, a significant fraction of the background seen in MiniBooNE was due to $\pi^{0}$ misidentification. This large background casts doubt on the significance of the excess seen, and so ways to deal with this background are necessary to further study the excess observed. MicroBooNE aims to investigate the excess seen in MiniBooNE by utilizing a Liquid Argon Time Projection Chamber to allow very high resolution particle tracking. This Bubble Chamber-like resolution will allow excellent background rejection, and hopefully resolve whether the excess seen in MiniBooNE is due to $\nu_{e}$ events or simply to a misunderstanding of the backgrounds. A simulated neutrino event in MicroBooNE can be seen in Figure \ref{fig:microboone}, where we can see the electron and proton tracks with high resolution. 

\subsection{MiniBooNE}

While MicroBooNE has taken data, MiniBooNE has continued to run. The original data set had $6.5\times10^{20}$ proton on target (POT). Since 2015, MiniBooNE has collected data on an additional $6\times10^{20}$ POT. 

Two analyses are currently planned with the new dataset. The first is to use identical procedure and cuts, and present separate results for the new runs and for a combined result. The second is to use new information that were not available with the original runs. 

\begin{figure}
\includegraphics[width = 0.5\textwidth]{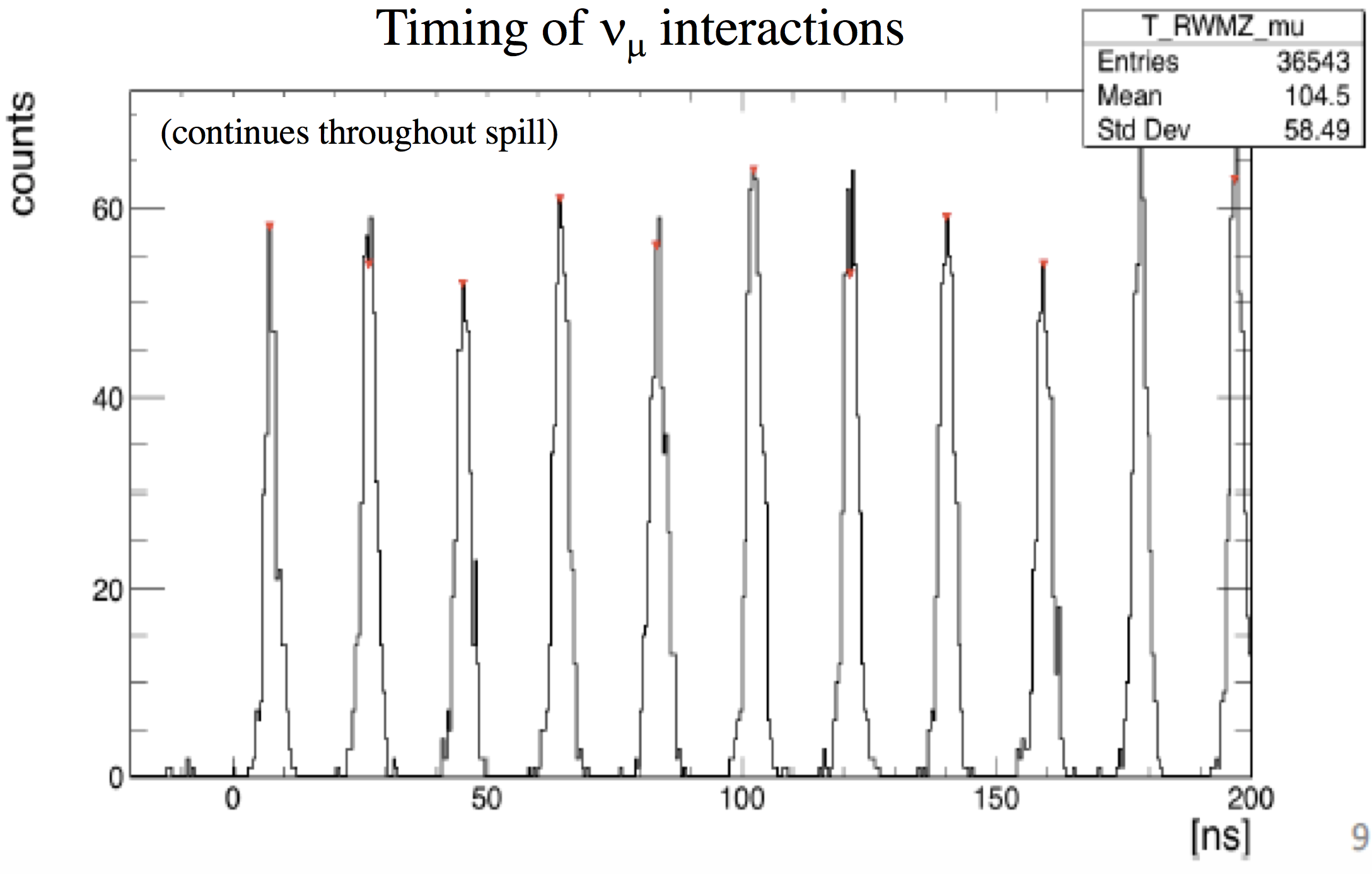}
\caption{}
\label{fig:bucket}
\end{figure}

One example of new information is beam-bucket timing, made possible by upgrades. The beam is delivered in 81 RF ``buckets", separated by 19.2 ns. MiniBooNE can determine the bucket maximum to within 1 ns, with a preliminary measured bucket width of 1.5 ns. Figure \ref{fig:bucket} demonstrates this accuracy. By requiring events to be in a bucket, we can cut on photon backgrounds. 

\bibliography{bibliography}

\end{document}